# Improving Grid Computing Performance by Optimally Reducing Checkpointing Effect

Aliyu Garba[a], A. F. D. Kana[a], Mohammed Abdullahi[a], Idris Abdulmumin[a], Shehu Adamu[b], Fatsuma Jauro[a]

{algarba, abdullahilwafu, iabdulmumin, fjaura}@abu.edu.ng, donfackkana@gmail.com, shehu.adamu@aun.edu.ng

[a]Department of Computer Science, Ahmadu Bello University, Zaria, Nigeria;
[b]SITC, American University of Nigeria, Yola, Nigeria

**Abstract:** Grid computing is a collection of computer resources that are gathered together from various areas to give computational resources such as storage, data or application services. This is to permit clients to access this huge measure of processing resources without the need to know where these might be found and what technology such as, hardware equipment and operating system was used. Dependability and performance are among the key difficulties faced in a grid computing environment. Various systems have been proposed in the literature to handle recouping from resource failure in Grid computing environment. One case of such system is checkpointing. Checkpointing is a system that endures faults when resources failed. Checkpointing method has the upside of lessening the work lost because of resource faults. However, checkpointing presents a huge runtime overhead. In this paper, we propose an improved checkpointing system to bring down runtime overhead. A replica is added to ensure the availability of resources. This replicates all checkpointing files to other machines as opposed to having dedicated checkpointing machine. The results of simulation employing GridSim noted that retaining the number of resources fixed and changing the number of gridlets, gains of up to 11%, 9%, and 11% on makespan, throughput, and turnaround time respectively, were realized while changing the number of resources and preserving the number of gridlets fixed, increases of up to 11%, 8%, and 9% on makespan, throughput, and turnaround time respectively, were realized.

**Introduction**

A Grid consists of resources that are loosely coupled and pervasively distributed to different geographical locations which are owned by different organizations (Jain and Chaudhary, 2014). Grid computing utilizes a computer network in which every computer is imparted to each other computer in the system. The objective of Grid is for computing to be widely spread, trustworthy, steady, low-cost such that individual clients can access computing resources such as processors, memory, data, and applications as required with deliberation to these clients. It is also meant to utilize accessible resources inside the grid to complete a process or application within the shortest conceivable time.

Several scheduling techniques based on nature-inspired algorithms such as Ant Colony Optimisation (ACO), Genetic Algorithm (GA), Cuckoo Search etc. have been proposed to achieve this goal. However, when some inevitable faults occur, they cause a delay in the speed of execution and thereby affect the performance (Garba *et al.*, 2017). Fault tolerance is a technique used to overcome the challenge of faults in the grid computing environment. Several approaches have been introduced to control the fault tolerance system to achieve high reliability and performance. Checkpointing is one of the techniques used to achieve high reliability. In this, a checkpoint is inserted at regular interval and if in the middle of an execution a failure occurs, the execution does not have to restart from the beginning but rather continue from the last checkpoint before the failure.

The major challenge with this technique is determining the optimal checkpoint interval as the high number of a checkpoint may exacerbate checkpoint overhead while in the event of a failure, a lower checkpoint may cause significant loss of execution. Besides, if a dedicated node is used to store the checkpoint file, the node is bound to fail like the other resources in the grid. Replication of checkpoint to other nodes within the grid will guarantee more reliability as compared to a dedicated checkpoint machine. This, though, will be at a cost of memory as well as reading and writing the checkpoint files. This work proposes to reduce checkpoint overhead and incorporate replica to improve reliability and performance as opposed to having dedicated checkpoint repository. Moreover, after every successive checkpoint, the checkpointing files are purged to free space.

*Checkpointing Techniques*

Checkpointing is a method that seeks to save the image of the status of processes at some regular point throughout its lifetime, and then, on the event of a failure, it can be continued at the point where the last checkpointed image is saved and successfully passed (Chen, Sun and Chen, 2016; Tran, Renault and Ha, 2017). Figure 1 illustrates how the checkpoint is invoked and replicated to the other node within the grid ecosystem to ensure reliability. After every successive checkpoint, say checkpoint 2, the previous checkpoint – checkpoint 1 – is purged out to free space, in that manner.

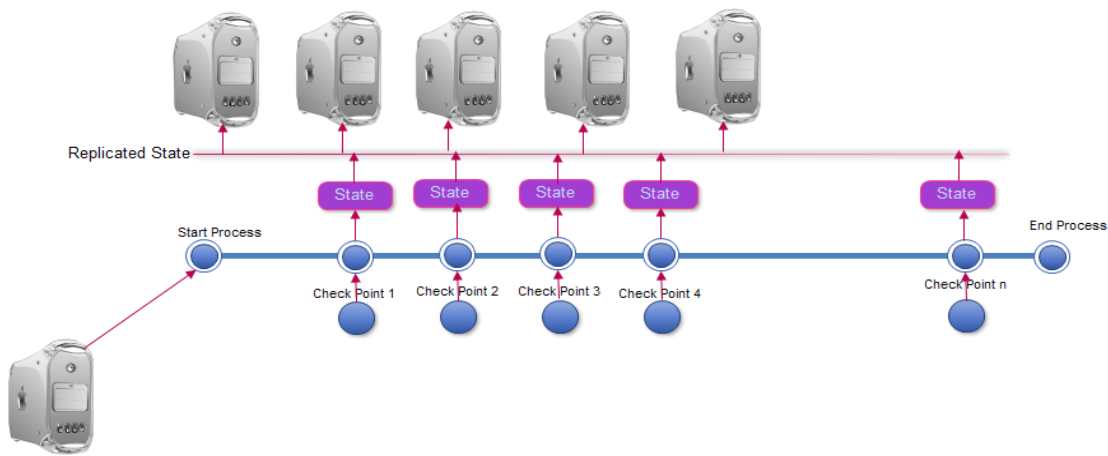

*Figure 1: Checkpointing Techniques*

Though different types of fault – physical, network, process, etc – can occur in real-time (Gupta, 2011), user-level checkpointing may not be the best for all circumstances. According to Meroufel and Belalem (2014), each of the checkpointing levels has its advantages as well as drawbacks in terms of flexibility, degrees of transparency, portability, restart-ability, efficiency, etc. However, based on the analysis, it is identified that mixed checkpointing levels provide better performance. Several kinds of checkpointing optimization (full or incremental checkpointing, unconditional periodic/optimal checkpointing, synchronous (coordinated) / asynchronous (uncoordinated) checkpointing, etc.) have been studied and analysed by researchers. Based on the analysis, an incremental checkpoint is the best technique (Agarwal *et al.*, 2004).

**Related Work**

The emergence of cloud and grid computing have attracted so many businesses and educational sectors in order to achieve high processing and reliable system within a minimum amount of time. This attracts researchers to focus on improving the performance in order to meet the demand for service vendors and their customers. Performance and reliability remain the principal aspects of all system. In Idris, Ezugu, Junaidu, and Adewumi (2017), fault-tolerant scheduling algorithms were embedded into an established ACO grid load balancing algorithm. The research was built on the work of Ludwig and Moallem (2011) and improvement was significantly made in terms of performance. However, this used a dedicated checkpoint repository for keeping and tracking the checkpointing files. In Sahadevan and Radhakrishnan (2014), the authors proposed a flexible job scheduling for an extension to the adaptation to internal failure that may occur using the checkpointing approach which could show a change in the execution of the general computational time even in most exceedingly terrible circumstance under heterogeneous environment. However, the study did not take into cognisance the size and number of checkpoints.

Several researchers (Balpande and Shrawankar, 2014; Idris *et al.*, 2017) have tried to propose techniques on how to reduce checkpointing overhead. Fault tolerance can be achieved at a different level of abstraction; User-level, System-level (Sancho *et al.*, 2005; Meroufel and Belalem, 2014) as briefly discussed in section 2. Moreover, the user-level provides better flexibility as it allows a programmer to invoke checkpoint as desired with complete control on where and when to. Sahadevan and Radhakrishnan (2014) employed the idea of checkpoint server where on every checkpoint set by the checkpoint manager, the job status is communicated to the checkpoint server.

The checkpoint server stores the job execution status and delivers it on demand during a resource failure. For an appropriate job, the checkpoint server drops the result of the preceding checkpoint meanwhile a new value of a checkpoint result is obtained and installed. User-level and programme-level are a sub-category of the application-level as illustrated and explained in Meroufel and Belalem (2014). It is observed that application-level has flexibility and transparency over the system-level (Meroufel and Belalem, 2014). The research conducted in Takizawa, Amrizal, Komatsu, and Egawa (2017), used the application-level with an incremental checkpointing mechanism to reduce the overhead caused by frequent checkpoint files. This used the concept of granularity with auto-tuning for reducing the checkpointing overhead.

This study tends to implement the proposed techniques suggested by Garba et al. (2017) in addition to combining the response time, failure rate and the tendency of resource failure in a bid to optimally reduce the checkpointing overhead and consequently improve the performance. The number of checkpoints is defined and decided before the execution of the job begins. In Addition, the replica is added to improve reliability.

**Methodology**

This section describes the improvement of the work by Idris et al. (2017) and the illustration on the process to be employed to realize the proposed checkpointing of the scheduling algorithms.

*Proposed Checkpointing Technique*

The architecture of the improved implementation checkpointing system is depicted in Figure 2.

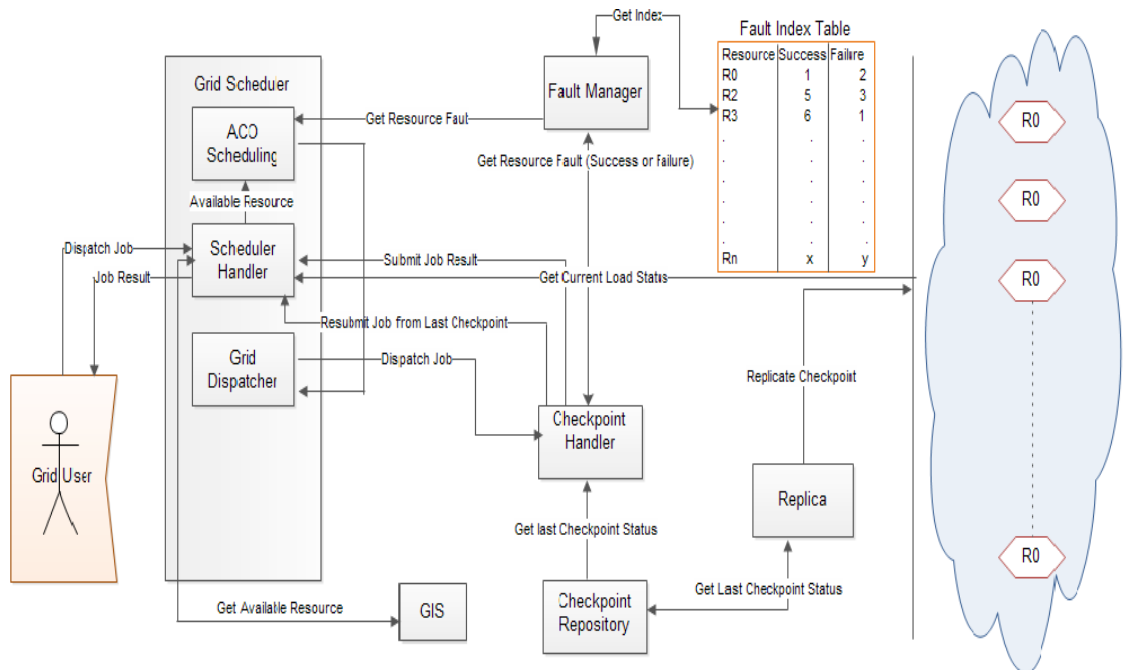

*Figure 2: Architecture of the fault tolerance in grid computing*

When a user submits a job to the scheduler handler, it communicates with GIS to receive a report about possible resources of the grid and then demands the resources to give their information about the most current workload state in which as well the fault manager transfers it fault rate history. A checkpoint point handler will determine the checkpoint interval based on the amount of failure rate, response time, and the tendency of failure of the failed resource. At each successive checkpoint, the replica is bound for replicating the execution states of the node to the other nodes of the grid to guarantee availability. If it does not come back alive, the states can be retrieved from the accessible one. If a failure occurs, the scheduler receives the set of suitable nodes with their workload and fault rate and it invokes the ACO techniques for decision making.

The ACO algorithm is in charge of looking through the resource that meets the client's requirement using fault rate and the workload on the resource before suitable decision making. The criteria for choosing a resource for scheduling depends on the present workload at hand and fault rate. A job dispatcher transmits the jobs to the checkpoint handler before sending to the job of the preferred resource. When a user presents the job to the dispatcher, the checkpoint handler communicates to the fault manager to know the number of triumphant jobs completed and the number of jobs failed on which it is scheduled and consequently sets the number of times required for the checkpoint and the appropriate interval.

When the failure rate is higher than the average failure rate, then there is a high possibility that the resource might not be admitted for Scheduling except that the workload of all other ones is doubled that of the failed one. Then the checkpoint handler submits the job with the checkpoints to the decided resource. The checkpoint handler stores the checkpoint interval to the checkpoint repository along with the unaccomplished executed result of the jobs which is obtainable in the checkpoint table. For a critical job, the checkpoint repository clears the replicated state of the preceding checkpoint to free space and recognizes the immediate triumphant checkpoint for recovery. The failure of the resource is notified to the checkpoint handler to increment its fault index. Contrarily, if a particular job is executed, it will be displaced from the checkpoint result table. The Fault handler tracks the state of the resources obtainable in the grid at constant intervals. The fault index provides the rate of resource failure. The higher the value of the fault index, the greater the failure rate of the resource but the lower failure rate if contrarily. Checkpoint handler suddenly asks the checkpoint repository to collect the most remiss checkpoint records of the accomplished jobs on the failed resource if possible or get it from the replicated checkpoint records and re-scheduled the jobs simultaneously with latest checkpoint state. On the victorious completion of the job, the checkpoint files are expelled to free space, and the checkpoint handler receives the job completion notification message from the grid and refreshes the fault index handler to add the success measure of the resource. The Fault Index Handler keeps a fault index records, which designates the failure rate of the resource.

*Checkpointing Algorithm*

The resolution to be used by the fault index handler on the update and preservation of the fault index is to be decided by the following modified algorithms as shown in Algorithms

1, 2, and 3. Algorithm 1 shows the modified portion of the scheduling algorithm of Idris *et al.*, (2017), where modifications are indicated in italic. Algorithm 2 and Algorithm 3 illustrate the existing and modified checkpoint algorithms respectively.

---

**Algorithm 1:** Modified Algorithm of the Fault Index

---

IF the checkpoint handler gets the job completion notification from the resource THEN

    Convey a report to the fault index handler to increment the success index of the resource.

    Submit completed job to the scheduler.

END IF

IF the checkpoint handler gets the failure report of the resource monitor THEN

    Forward a report to the fault index handler to increment the failure index of the resource that fails to accomplish the designated job.

    Forward a report to the checkpoint repository to check the checkpoint state of this job.

    *Replicate the checkpointing state to the other resources.*

    IF the result in the checkpoint repository of a job exits THEN

        Relinquish the immediate checkpoint data obtained to the scheduler handler for re-scheduling.

        Exit

    *ELSE*

        *Get the replicated checkpoint data from any available resource.*

        *Submit the retrieved replicated data files to the scheduler handler for rescheduling*

        *Exit*

    END IF

END IF

---

**Algorithm 2:** Existing Checkpoint Algorithm

---

    S1: set checkpoint time and interval for a particular job

    S2: submit the job for execution

        While job is running till the end

            S3: IF currentTime ≥ checkpointTime THEN

                Update checkpoint interval time

                Increment number of time checkpoint

                Checkpoint current process state

            Else not time for checkpoint

**Algorithm 3:** Modified Checkpoint Algorithm

    P0: failure rate of particular resource
    P1: tendency of failure of a particular resource
    P2: the average number of all resources failed
    IF (P1>P2) THEN
        P3: set checkpoint time and the interval to be high
    ELSE IF (P0>=P2) THEN // there is high failure rate
        P3: set checkpoint time and the interval to be high
        P4: present the job for processing
    While job is processing till the end
        IF current_time = checkpoint_time THEN
            Set update to the checkpoint time interval
            Increment the number of times for the checkpoint
            Store state current process
        Else
            It is not the time for checkpoint

*System Implementation*

The checkpointing is one of the most common general and extensively used method to provide fault-tolerance on unstable systems. It is a record of the snapshot of the whole system state to restart the application following the existence of any failure. The states and data of an executing job involve registers comprising the data structures, address, memory spaces, libraries, and files containing data with big size. A programming-level checkpointing control is assumed.

        The resource failure rate (FR) and the tendency of failure are used to define the checkpoint interval and the number of checkpoints alternative to using the resource fault index. The resource failure rate is used to describe the failure history of the resource. The mathematical models used to measure the number of a checkpoint and the optimal checkpoint interval are detailed below; equation 1 determines the fault rates of the resources and equation 2 estimates the number of times to checkpoint while it is executing and equation 3 estimates the checkpoint interval of time and when a checkpoint should be transpired.

$$Failure\ Rate(FR) = F_b/N_b \qquad (1)$$

        where $F_b$ is the resource that failed during execution and the $N_b$ is the number of jobs submitted to the grid.

Suppose $R_t$ is the response time, then the number of checkpoints can be determined by:

$$Number\ of\ Checkpoints\ (C_n) = R_t * FR \qquad (2)$$

The interval between one checkpoint and the other (checkpoint interval) is given by:

$$Checkpoint\ Interval = R_t/C_n \qquad (3)$$

$$Fault\ Tendency(FD) = \frac{\sum_{j=1}^{n} p_{fj}}{n} * 100\% \qquad (4)$$

where $n$ is the total number of grid resources greater than zero and $p_{fj}$ is the failure rate of resource $j$. Through this metric, we can foresee the fault behaviour of the system (Altameem, 2013).

**Results and Discussion**

The results of this finding are compared with Idris et al. (2017) using the same system requirements – shown in Table 1 and Table 2 below – and under the same heterogeneous environment. The outcomes of the research are analysed based on the performance metrics; makespan, throughput, and turnaround time.

| No. of Machines/Resource | 1 |
|---|---|
| No. of Processing Elements (PE)/Machine | 2 |
| PE Ratings | 50 MIPS |
| Bandwidth | 5000 B/S |

*Table 1: Grid Resource Characteristics*

| Length | 0 – 50000 MI |
|---|---|
| Input File Size | 100 + (10% to 40%) |
| Output File Size | 250 + (10% to 50%) |

*Table 2: Gridlet Characteristics*

In the simulation, the scheduling analyses are conducted by retaining the number of resources fixed and setting different values to the number of jobs; the number of gridlet varied from 1,00 to 3,700 while the length of per gridlet is 200000MIs at every step and examined as performed in Idris et al. (2017) for comparative analysis.

To evaluate and contrast the achievement of the suggested checkpoint algorithm, throughput (T) is one of the most extensive performance metrics that was used to measure the performance of fault-tolerant systems and is mathematically defined as:

$$\text{Throughput }(T) = N_j/T_n \qquad (5)$$

where $N_j$ is the number of jobs submitted to the Grid by user and $T_n$ is the total amount of time it expects to finish the execution of n jobs.

Throughput estimates the number of batch jobs executed in a certain period, Rise in throughput in the proposed checkpoint algorithm implied the improvement by about 9%. Table 3 shows the improvement realized based on throughput. The tendency of failure perceived during simulation reveals that the number of jobs failure rises as the number of Gridlets increased. This has to do with resources that have more exceeding jobs in execution at the time of failure. Consequently, there is more improvement in throughput as the number of gridlets increases.

| No. of Resources | No. of Gridlets | Existing Throughput (ms) | Improved Throughput (ms) | Improvement (%) |
|---|---|---|---|---|
| 100 | 100 | 0.001296 | 0.001322 | 1.97 |
| 100 | 500 | 0.00322 | 0.003381 | 4.76 |
| 100 | 900 | 0.003909 | 0.004104 | 4.75 |
| 100 | 1300 | 0.004431 | 0.004697 | 5.66 |
| 100 | 1700 | 0.004955 | 0.005253 | 5.67 |
| 100 | 2100 | 0.005526 | 0.005912 | 6.53 |
| 100 | 2500 | 0.005947 | 0.006363 | 6.54 |
| 100 | 2900 | 0.006234 | 0.006671 | 6.55 |
| 100 | 3300 | 0.006553 | 0.007077 | 7.40 |
| 100 | 3700 | 0.006762 | 0.00737 | 8.25 |

*Table 3: Average Throughput Table for Varied Gridlets*

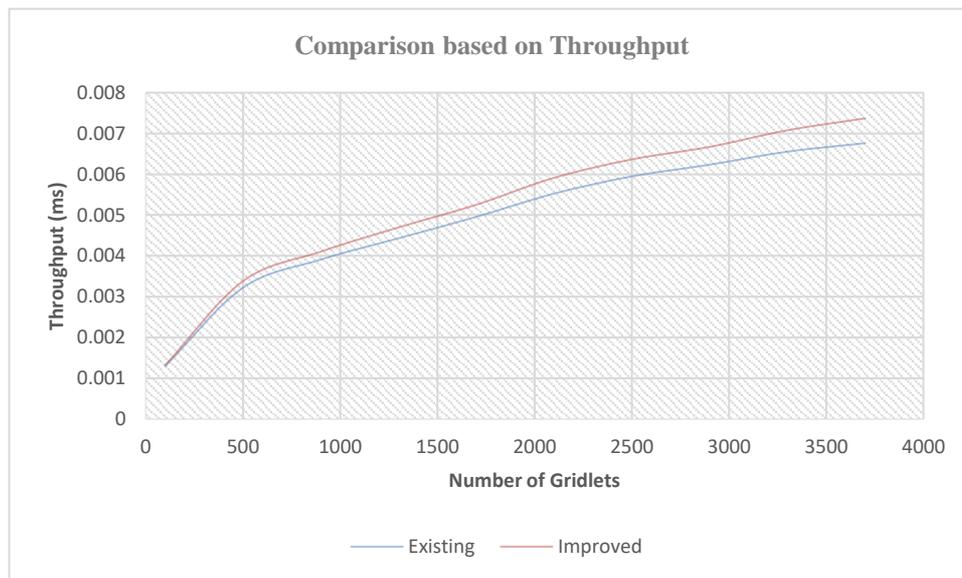

*Figure 3: Graph of Average Throughput for varied Gridlets*

Figure 3 shows the aggregate comparative achievement against throughput within the two systems over changing numbers of gridlets. From the results, despite the throughput enhancement, we observe the effect of the overhead of checkpoint management in our proposed model. [Figure 3 near here]

Results of the analysis indicated that there is a reduction in makespan as compared with Idris et al. (2017). This showed an improvement of about 11% as illustrated in Table 4 below. This is simply because the checkpoint is not frequently inserted that could bring an overhead and consequently increased the job execution time. Figure 4 shows the aggregate comparative analysis of makespan within the two systems over changing numbers of gridlets.

| No. of Resources | No. of Gridlets | Existing Makespan | Improved Makespan | Improvement (%) |
|---|---|---|---|---|
| 100 | 100 | 82581.19 | 78452.56 | 5.26 |
| 100 | 500 | 159281.1 | 148131.42 | 7.53 |
| 100 | 900 | 235238.1 | 218771.46 | 7.53 |
| 100 | 1300 | 298183.8 | 274329.11 | 8.70 |
| 100 | 1700 | 347092.4 | 315854.06 | 9.89 |
| 100 | 2100 | 383579.5 | 352893.12 | 8.70 |
| 100 | 2500 | 423205.2 | 385116.73 | 9.89 |
| 100 | 2900 | 467180.2 | 420462.18 | 11.11 |
| 100 | 3300 | 505965 | 470547.41 | 7.53 |
| 100 | 3700 | 548524.3 | 488186.66 | 12.36 |

*Table 4: Average Makespan for Varied Gridlets*

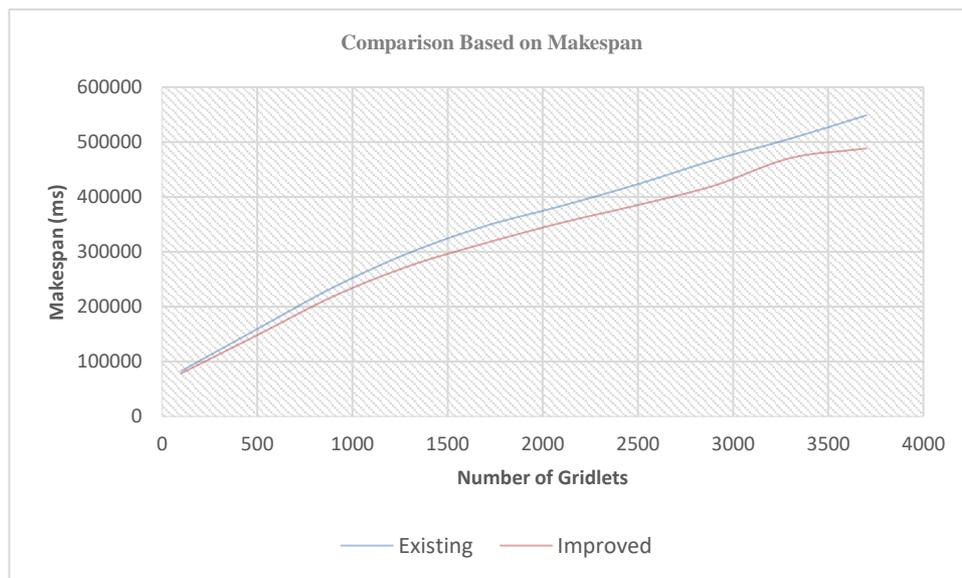

*Figure 4: Graph of average Makespan for varied Gridlets*

The size of the Gridlets advanced the average turnaround time by 11% as shown in Table 5. The increase was a result of the occasional insertion of the checkpoint and the selection of resources by examining the estimated low tendency of failure. Figure 5 presents the aggregate comparative analysis on average ATAT within the two methods over changing numbers of gridlets.

| No. of Resources | No. of Gridlets | Existing ATAT | Improved ATAT | Improvement (%) |
|---|---|---|---|---|
| 100 | 100 | 32851.02 | 31865.49 | 3.09 |
| 100 | 500 | 79757.59 | 75769.71 | 5.26 |
| 100 | 900 | 134389.9 | 126326.51 | 6.38 |
| 100 | 1300 | 186865.8 | 171916.58 | 8.70 |
| 100 | 1700 | 228997.5 | 215257.66 | 6.38 |
| 100 | 2100 | 262358.6 | 241369.92 | 8.70 |
| 100 | 2500 | 298553.1 | 271683.33 | 9.89 |
| 100 | 2900 | 339709.5 | 315929.85 | 7.53 |
| 100 | 3300 | 383153.9 | 341007.01 | 12.36 |
| 100 | 3700 | 427488.4 | 384739.57 | 11.11 |

*Table 5: Average Turnaround Time for Varied Gridlet*

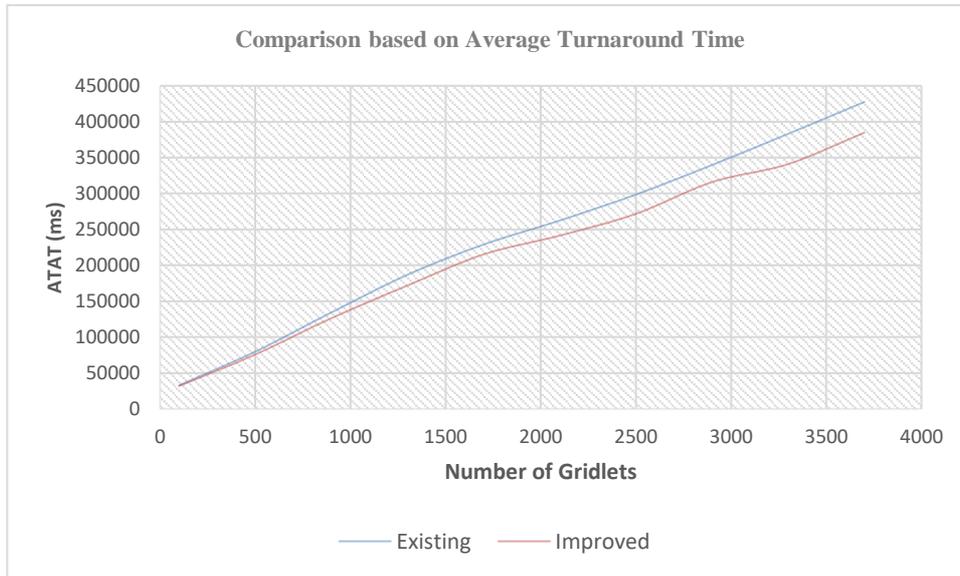

*Figure 5: Graph of Average Turnaround Time for varied Gridlets*

Now retaining the number of Gridlets fixed and changing the number of resources, the outcomes show that there are gains in terms of throughput, makespan and average turnaround time by up to 8%, 11% and 9% as shown in Table 6, Table 7 and Table 8 respectively. Figure 6, Figure 7, and Figure 8 illustrate the aggregate comparative analysis against throughput, makespan, and ATAT, respectively within the two systems over varying numbers of jobs.

| No. of Gridlets | No. of Resources | Existing Throughput (ms) | Improved Throughput (ms) | Improvement (%) |
|---|---|---|---|---|
| 3000 | 50 | 0.000723 | 0.0007594 | 4.79 |
| 3000 | 250 | 0.003297 | 0.0034573 | 4.64 |
| 3000 | 450 | 0.00477 | 0.0050561 | 5.66 |
| 3000 | 650 | 0.005824 | 0.0061155 | 4.77 |
| 3000 | 850 | 0.006571 | 0.0069653 | 5.66 |
| 3000 | 1050 | 0.007189 | 0.0075485 | 4.76 |
| 3000 | 1250 | 0.007919 | 0.0083937 | 5.66 |
| 3000 | 1450 | 0.008372 | 0.00879 | 4.76 |
| 3000 | 1650 | 0.008859 | 0.0095678 | 7.41 |
| 3000 | 1850 | 0.009327 | 0.0099796 | 6.54 |
| 3000 | 2050 | 0.009683 | 0.0104571 | 7.40 |
| 3000 | 2250 | 0.009845 | 0.0102387 | 3.85 |
| 3000 | 2450 | 0.010471 | 0.0108894 | 3.84 |
| 3000 | 2650 | 0.010837 | 0.0114867 | 5.66 |
| 3000 | 2850 | 0.010826 | 0.0111512 | 2.92 |
| 3000 | 3050 | 0.011195 | 0.0120915 | 7.41 |

*Table 6: Average Throughput Time for Varied Resources*

| No. of Gridlets | No. of Resources | Existing Makespan | Improved Makespan | Improvement (%) |
|---|---|---|---|---|
| 3000 | 50 | 4193551.12 | 3858067.03 | 8.70 |
| 3000 | 250 | 915672.954 | 833262.39 | 9.89 |
| 3000 | 450 | 635026.91 | 584224.76 | 8.70 |
| 3000 | 650 | 522718.722 | 486128.41 | 7.53 |
| 3000 | 850 | 462654.445 | 430268.63 | 7.53 |
| 3000 | 1050 | 425486.675 | 391447.74 | 8.70 |
| 3000 | 1250 | 384175.14 | 349599.38 | 9.89 |
| 3000 | 1450 | 366323.478 | 337017.61 | 8.70 |
| 3000 | 1650 | 324779.388 | 292301.45 | 11.11 |
| 3000 | 1850 | 330142.321 | 300429.51 | 9.89 |
| 3000 | 2050 | 317569.953 | 292164.36 | 8.70 |
| 3000 | 2250 | 311514.311 | 293478.02 | 6.15 |
| 3000 | 2450 | 293752.861 | 264377.57 | 11.11 |
| 3000 | 2650 | 282355.198 | 251296.13 | 12.36 |
| 3000 | 2850 | 284203.798 | 261467.51 | 8.70 |
| 3000 | 3050 | 272435.062 | 247915.92 | 9.89 |

*Table 7: Average Makespan Time for Varied Resource*

| No. of Gridlets | No. of Resources | Existing ATAT | Improved ATAT | Improvement (%) |
|---|---|---|---|---|
| 3000 | 50 | 3026939.18 | 2784784.04 | 8.70 |
| 3000 | 250 | 569764.08 | 535578.23 | 6.38 |
| 3000 | 450 | 324526.24 | 301809.41 | 7.53 |
| 3000 | 650 | 232252.33 | 213672.15 | 8.70 |
| 3000 | 850 | 186910.56 | 170088.62 | 9.89 |
| 3000 | 1050 | 159492.9 | 148328.39 | 7.53 |
| 3000 | 1250 | 141939.98 | 140520.58 | 1.01 |
| 3000 | 1450 | 130883.26 | 120412.6 | 8.70 |
| 3000 | 1650 | 121695.2 | 111959.59 | 8.70 |
| 3000 | 1850 | 116649.38 | 106150.93 | 9.89 |
| 3000 | 2050 | 112504.84 | 103504.45 | 8.70 |
| 3000 | 2250 | 109215.73 | 100478.47 | 8.70 |
| 3000 | 2450 | 107411.06 | 100966.4 | 6.38 |
| 3000 | 2650 | 104947.09 | 99699.74 | 5.26 |
| 3000 | 2850 | 103099.57 | 95882.5 | 7.53 |
| 3000 | 3050 | 101897.4 | 93745.61 | 8.70 |

*Table 8: Average Throughput Time for Varied Resources*

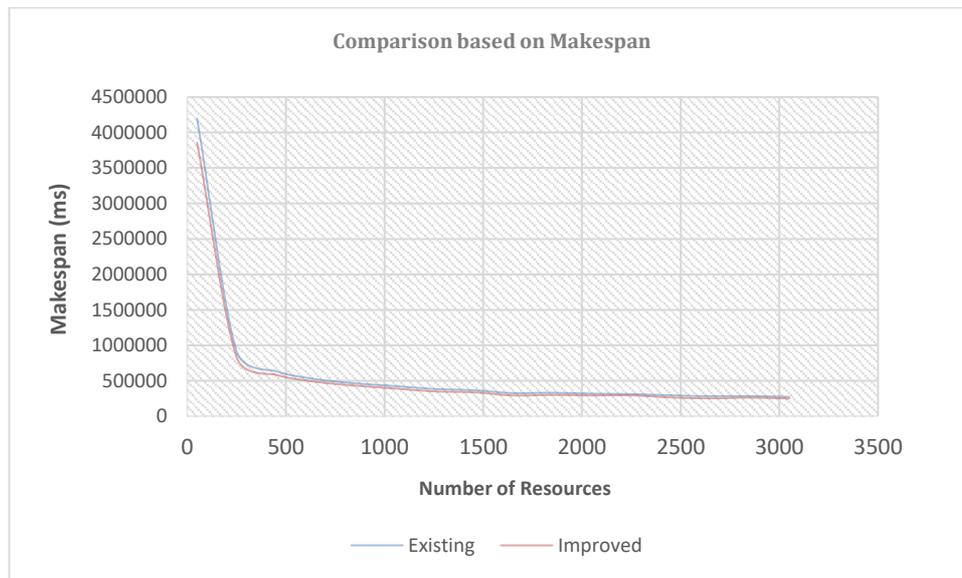

*Figure 6: Graph of Average Makespan time for varied Number of Resources*

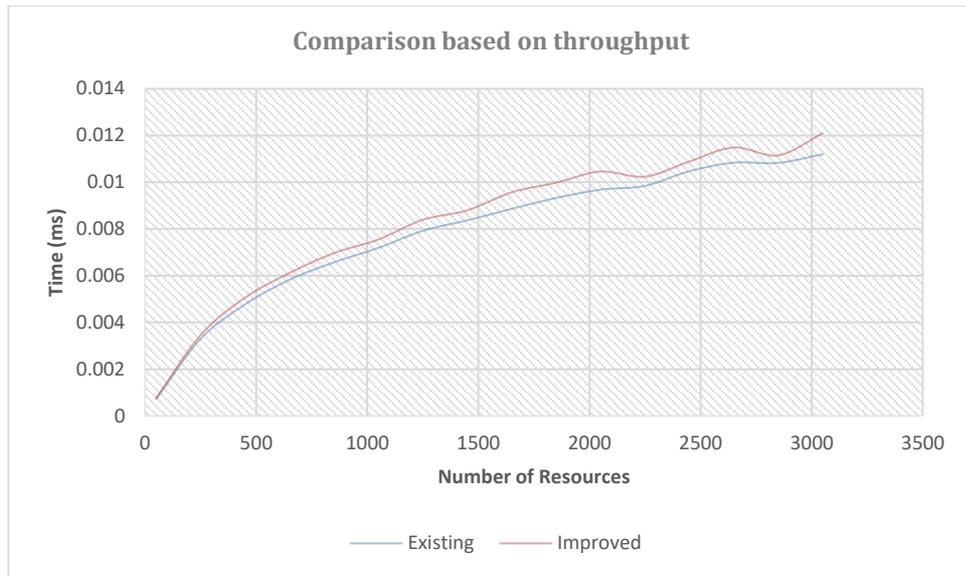
*Figure 7: Graph Throughput for varied Gridlets*

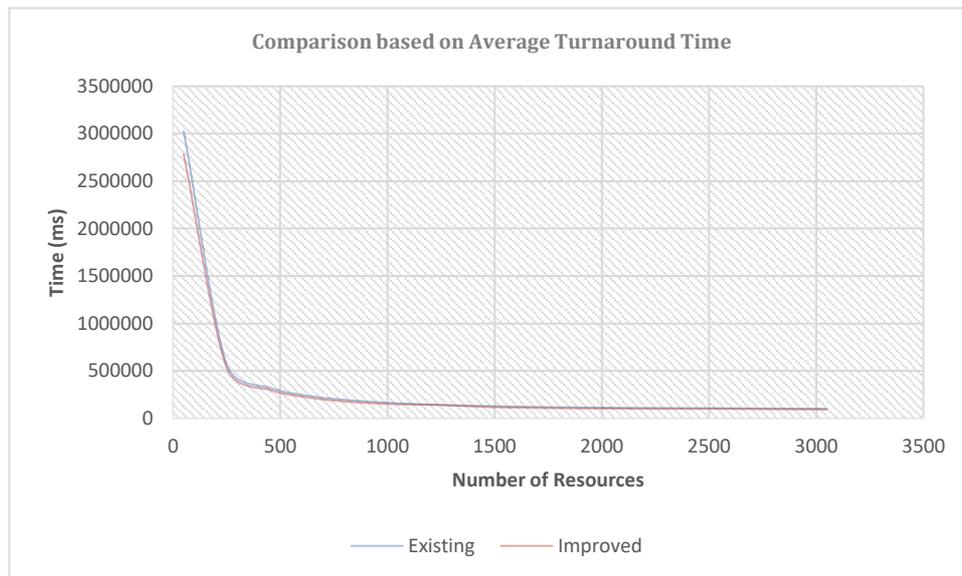
*Figure 8: Graph of Average Turnaround time for varied Number of resources*

**Conclusion and Future Work**

The fault-tolerance method was applied to address the issue of fault whereby the checkpointing mechanism was adopted to avoid restarting the execution of the user's application, thereby enhancing performance. Consequently, producing many checkpoints result in overheads which affect the performance. This study focused on the issue of overheads associated with many checkpointing. The result of the simulation obtained using the GridSim simulator revealed modest performance improvements in terms of makespan, throughput, and turnaround time. The major contributions to this study are:

- the restoration approaches are achieved by consolidating checkpointing and replication,
- the expected number of checkpoints is settled earlier to the job's execution,
- the checkpoint size is based on the response time, failure rate and tendency of failure, and
- the checkpoint file is cleared following every succeeding checkpoint.

One area for future work is the issue of allocation and deallocation of memory; the checkpoint files are replicated to all other nodes in the grid alongside the checkpoint repository, resulting to provoking the expense of keeping and reclaiming the checkpoint files.